\documentclass[aps, prl, twocolumn,showkeys, showpacs,amsmath,amssymb,superscriptaddress]{revtex4}
\usepackage{graphicx}
\usepackage{dcolumn}
\usepackage{bm}
\usepackage{epsfig}

\begin{document}
\title{Lattice Boltzmann scheme for relativistic fluids}

\author{M. Mendoza} \email{mmendoza@ethz.ch} 
\affiliation{ ETH Z\"urich, Computational Physics for Engineering Materials,
  Institute for Building Materials, Schafmattstrasse 6, HIF, CH-8093 Z\"urich
  (Switzerland)}

\author{B. Boghosian}\email{bruce.boghosian@tufts.edu}
\affiliation{Bromfield-Pearson, Medford, Massachusetts 02155,
  Department of Mathematics, Tufts University }

\author{H. J. Herrmann}\email{hjherrmann@ethz.ch}
\affiliation{ ETH Z\"urich, Computational Physics for Engineering Materials,
  Institute for Building Materials, Schafmattstrasse 6, HIF, CH-8093 Z\"urich
  (Switzerland)}

\author{S. Succi} \email{sauro.succi@gmail.com} 
\affiliation{Istituto per le Applicazioni del Calcolo C.N.R., Via dei Taurini, 19 00185,
  Rome (Italy),\\and
Freiburg Institute for Advanced Studies, Albertstrasse, 19, D-79104, Freiburg,
Germany}

\date{\today}
\begin{abstract}
  A Lattice Boltzmann formulation for relativistic fluids is presented
  and numerically verified through quantitative comparison with recent
  hydrodynamic simulations of relativistic shock-wave propagation in
  viscous quark-gluon plasmas.  This formulation opens up the
  possibility of exporting the main advantages of Lattice Boltzmann
  methods to the relativistic context, which seems particularly useful
  for the simulation of relativistic fluids in complicated geometries.
\end{abstract}



\maketitle

%Relativistic fluid dynamics plays an important role in different
%fields like, e.g.  astrophysics and high energy physics. However, the
%dynamics of such systems involves solving highly nonlinear equations
%and so analytical solutions of practical problems cannot be easily
%found.  Therefore, several numerical methods have been developed based
%on macroscopic continuum description \cite{NumMethod2, NumMethod4,
%  NumMethod5} and kinetic theory\cite{NumMethod1}.
In the last decade, the Lattice Boltzmann (LB) method has attracted
considerable interest as an alternative computational fluid dynamics
method, based on the solution of a minimal Boltzmann kinetic equation,
rather than on the discretization of the equations of continuum fluid
mechanics \cite{LBE1, LBE2}.  To date, the overwhelming majority of LB
applications are directed towards classical, i.e. non-quantum and
non-relativistic, fluids.  However, while quantum versions of the LB
equation have existed for more than a decade \cite{QLB}, to the best
of our knowledge, an LB equation capable of handling relativistic
fluids has not yet been proposed. In this Letter, we fill this gap and
present an LB formulation for relativistic fluids.  Our procedure is
based on two simple and yet apparently unpursued observations, i) the
kinetic formalism is naturally covariant/hyperbolic/conservative, ii)
being based by construction on a finite-velocity scheme, existing
lattice Boltzmann methods naturally feature relativistic-like
equations of state, in the sense that the sound speed, $c_s$, is a
sizeable fraction of the speed of light $c$, i.e. the maximum velocity
of mass transport ($c_s/c=K$, with $0.1<K<1$).  Based on the above,
one is led to propose that, upon choosing the lattice speed
$c_l$$\equiv$$\delta x /\delta t \sim c$, the current LB mathematical
framework should allow for relativistic extensions, which is indeed
the case as shown in this Letter.

\begin{figure}[h!]
  \begin{center}
    \includegraphics[trim = 95mm 56mm 80mm 35mm, clip,
    width=8.2cm, height=5cm]{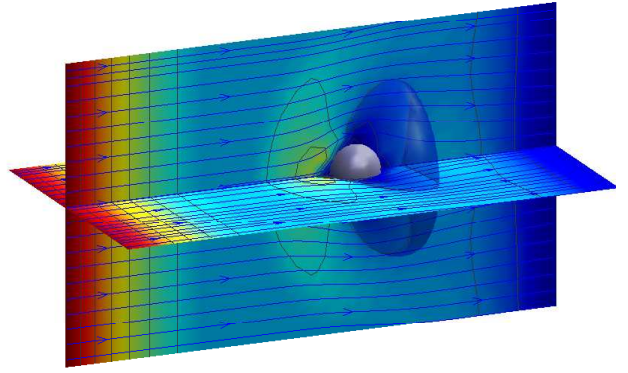}
    \caption{Relativistic shock wave impacting on a sphere at
      $|\vec{\beta}|=0.5$.  Here the streamlines represent the
      velocity field, and the colors the pressure. The simulation was
      implemented using a grid size of $200 \times 100 \times 100$
      cells.}\label{picture3D}
  \end{center}
\end{figure}
Our relativistic LB scheme (RLB) relies upon a moment-matching
procedure similar to the one originally used to derive the earliest LB
models for classical hydrodynamics. That is, the local kinetic
equilibria are expressed as parametric polynomials of the relativistic
fluid velocity $\vec{\beta} = \vec{u}/c$, with the parameters fixed by
the condition of matching the analytic expression of the relevant
relativistic moments, namely the number density, energy density and
energy-momentum.  As anticipated, the possibility of a successful
matching stems directly from the fact that, even in standard
(non-relativistic) LB fluids, the sound speed $c_s$ is of the same
order of the speed of light, typically $c_s=c/\sqrt 3$, which is
exactly the equation of state of relativistic fluids.  As a result,
$|\vec{\beta}| = Ma/\sqrt 3$, so that $|\vec{\beta}|$ is of the same
order as the Mach number $Ma=|\vec{u}|/c_s$. Owing to this simple, and
yet crucial property, it is therefore possible to tackle weakly
relativistic problems much the same way as traditional LB handles
classical low-Mach fluids.  This spawns the exciting opportunity of
carrying the assets of LB over to the context of weakly relativistic
fluids, such as the important case of quark-gluon plasmas generated by
recent experiments on heavy-ions and hadron jets \cite{QG-1, QG-2,
  QG-3, QG-4, QG-5, QG-6, QG-7}.  The RLB scheme is verified through
quantitative comparison with recent one dimensional hydrodynamic
simulations of relativistic shock-wave propagation in viscous
quark-gluon plasmas \cite{BAMPSs}. We can also apply our scheme to
three dimensional geometries as shown in Fig.  \ref{picture3D}.

Being based on a second-order moment-matching procedure, rather than
on a high-order systematic expansion in $\vec{\beta}$ of the local
relativistic equilibrium (J\"uttner) distribution\cite{NumMethod1},
the present approach is in principle limited to weakly relativistic
problems, with $|\vec{\beta}| \sim 0.1$.  However, by introducing
artificial faster-than-light particles (numerical ``tachyons''), the
present RLB scheme is shown to produce quantitatively correct results
up to $|\vec{\beta}| \sim 0.6$, corresponding to Lorentz's factors
$\gamma =\frac{1}{\sqrt{1-|\vec{\beta}|^2}} \sim 1.4$.  Although still
far from state-of-the-art numerical methods for relativistic
hydrodynamics \cite{NUMREL}, the RLB might nevertheless offer a fairly
inexpensive alternative to more sophisticated methods at moderate
values of $|\vec{\beta}|$.  In addition, since LB is recognizedly an
excellent solver for flows in complex geometries, like porous media,
it is plausible to expect that the present RLB scheme may play a
useful role for the simulation of relativistic fluids in non-idealized
geometries.

To begin our model description, we focus on the relativistic fluid
equations associated with the conservation of number of particles and
momentum-energy. The energy-momentum tensor reads as
follows\cite{RelaBoltEqua, paperRomat}: $T^{\mu \nu}= P \eta^{\mu \nu}
+ (\epsilon +P)u^\mu u^\nu$, with $\epsilon$ the energy density and
$P$ the hydrostatic pressure. The velocity 4-vector is defined by
$u^{\mu}= (\gamma, \gamma \vec{\beta})^{\mu}$ , where
$\vec{\beta}=\vec{u}/c$ is the velocity of the fluid in units of the
light speed and $\gamma$$=$$\frac{1}{\sqrt{1-|\vec{\beta}|^2}}$.  The
tensor $\eta^{\mu \nu}$ denotes the Minkowski metric.  Additionally,
we define the particle 4-flow, $N^{\mu}= (\gamma n, n \gamma
\vec{\beta})^{\mu}$, with $n$ the number of particles per volume.
Applying the conservation rule to energy and momentum, $\partial_\mu
T^{\mu \nu} = 0$, and to the 4-flow, $\partial_\mu N^{\mu} = 0$, we
obtain the hydrodynamic equations.
%\begin{subequations}\label{macroeq1}
%  \begin{equation}
%    \frac{\partial}{\partial t}\left( (\epsilon+P)\gamma^2-P \right) + \nabla
 %   \cdot \left((\epsilon + P)\gamma^2 \vec{u} \right) = 0 \quad ,
%  \end{equation}
%  \begin{equation}
%    \frac{\partial}{\partial t}\left( (\epsilon+P)\gamma^2 \vec{u} \right) +
%    \nabla P + \nabla \cdot \left( (\epsilon + P)\gamma^2 \vec{u} \vec{u}
%    \right) = 0 \quad ,
%  \end{equation}
%\end{subequations}
%for the energy mometum conservation, and
%\begin{equation}\label{macroeq2}
%    \frac{\partial n \gamma}{\partial t} + \nabla \cdot \left( n\gamma \vec{u}
%    \right) = 0 \quad ,
%\end{equation}
%for the conservation of particle number.  

Note that as opposed to a non-relativistic fluid, we have two scalar
equations. To complete the set of equations, we need to define a state
equation that relates, at least, two of the three quantities: $n$, $P$
and $\epsilon$.

The above hydrodynamic equations can be derived as a macroscopic limit
of the following relativistic Boltzmann-BGK equation
\cite{RelaBoltEqua, paperRomat}
\begin{equation}
  \label{RBGK}
  \partial_{\mu} (p^{\mu} f) = \frac{f^{eq}-f}{c \tau} 
\end{equation}
where $p^{\mu} = (E(p), \vec{p}c)$ is the particle 4-momentum with
$E(p)$ the relativistic energy as function of the momentum modulus
$p$$=$$|\vec{p}|$, $f^{eq}$ a local relativistic equilibrum and $\tau$
the relaxation time.  Lattice Boltzmann theory for classical fluids
shows that it may prove more convenient to solve fluid problems by
numerically integrating the underlying kinetic equation rather than
the macroscopic fluid equation themselves. The main condition for this
to happen is that a sufficiently economic representation of the
velocity space degrees of freedom be available.  Following upon
consolidated experience with non-relativistic fluids, such a
representation is indeed provided by discrete lattices, whereby the
particle velocity (momentum) is constrained to a handful of constant
discrete velocities, with sufficient symmetry to secure the
fundamental conservations of fluid flows, namely mass-momentum-energy
conservation and rotational invariance.  The main advantages of the
kinetic representation for classical fluids have been discussed at
length\cite{LBE2}, and they amount basically to the fact that the
information is transported along straight-streamlines (the discrete
velocities are constant in space and time) rather than along
space-time dependent trajectories generated by the flow itself, as it
is case for hydrodynamic equations.  Moreover, diffusive transport is
not described by second-order spatial derivatives, but rather emerges
as a collective property from adiabatic relaxation to local
equilibria.  This is crucial in securing a balance between first-order
derivatives in both space and time, which is essential for
relativistic equations.
%These advantages are expected to become even more
%visible in moving from non-relativistic to relativistic fluids, where
%hyperbolicity is deeply connected to relativistic covariance and
%finite-speed Maxwell-Cattaneo causality.

\begin{figure}
  \centering \includegraphics[scale=0.4]{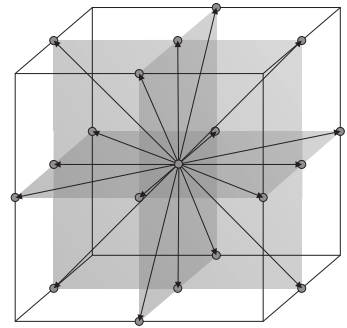}
  \caption{Set of discrete velocities for the relativistic lattice
    Boltzmann model. The highest speed is $\sqrt{2}c_l$}\label{d3q19}
\end{figure}

In order to reproduce the relativistic hydrodynamic equations, we
propose an LB model with the D3Q19 cell configuration, as shown in
Fig.  \ref{d3q19}. We define two distribution functions $f_i$ and
$g_i$ for each velocity vector $\vec{c}_{i}$, where the index $i$
labels the discrete momenta within each cell. The hydrodynamic
variables are calculated by imposing the following macroscopic
constraints, $n\gamma=\sum_{i=0}^{18} f_{i}$, $(\epsilon + P)\gamma^2
- P= \sum_{i=0}^{18} g_{i}$, and $(\epsilon + P)\gamma^2
\vec{u}=\sum_{i=0}^{18} g_{i} \vec{c}_{i}$. From these equations, we
have to extract the physical quantities $n$, $\vec{u}$, $\epsilon$ and
$P$, where we have only five equations for six unknowns. The problem
is closed by choosing an equation of state for ultra-relativistic
fluids, $\epsilon$$=$$3P$.

The distribution functions evolve according to the BGK Boltzmann
evolution equation \cite{BGK} (full details in a future extended
publication),
\begin{eqnarray}{\label{lbe1}}
  f_{i}(\vec{x}+\vec{c}_i \delta t,t+\delta
  t)-f_{i}(\vec{x},t)&=-\frac{\delta t}{\tau} (f_i - f_i^{\rm eq}) \quad ,
\end{eqnarray} 
and,
\begin{eqnarray}{\label{lbe2}}
  g_{i}(\vec{x}+\vec{c}_i\delta t,t+\delta
  t)-g_{i}(\vec{x},t)&=-\frac{\delta t}{\tau} (g_i - g_i^{\rm eq}) \quad ,
\end{eqnarray}
where $f_i^{\rm eq}$ and $g_i^{\rm eq}$ are the equilibrium
distribution functions.

The equilibrium distribution functions that recover the relativistic
fluid equations in the continuum limit, read as follows:
\begin{eqnarray}{\label{equil1}}
  f_i^{\rm eq} = w_i n \gamma \left[1+3\frac{(\vec{c}_i \cdot \vec{u})}{c_l^2} \right] \quad ,
\end{eqnarray} 
for $i$$\ge$$0$ and, 
\begin{eqnarray}{\label{equil2a}}
  \begin{aligned}
    g_i^{\rm eq} = w_i (\epsilon +P) \gamma^2 \biggl[\frac{3
      P}{(P+\epsilon)\gamma^2 c_l^2} +3\frac{(\vec{c}_i \cdot
      \vec{u})}{c_l^2} \\ + \frac{9}{2}\frac{(\vec{c}_i \cdot
      \vec{u})^2}{c_l^4} - \frac{3}{2}\frac{|\vec{u}|^2}{c_l^2}
    \biggr] \quad ,
  \end{aligned}
\end{eqnarray} 
for $i$$>$$0$ and,
\begin{eqnarray}{\label{equil2b}}
  g_0^{\rm eq} = w_0 (\epsilon +P) \gamma^2 \left[ 3 - \frac{3 P
      (2+c_l^2)}{(P+\epsilon)\gamma^2 c_l^2} -
    \frac{3}{2}\frac{|\vec{u}|^2}{c_l^2} \right] \quad ,
\end{eqnarray} 
for the rest particles. Here, $c_l$ is the limiting velocity of the
lattice which relates the cell size and the time step
$c_l$$=$$\frac{\delta x}{\delta t}$, and we have rescaled the velocity
units such that the speed of light $c$$=$$1$. The weights for this set
of discrete speeds are defined by $w_{0}=1/3$ for the rest particles,
$w_i=1/18$ for the velocities $|\vec{c}_i|$$=$$c_l$, and $w_i=1/36$
for $|\vec{c}_i|$$=$$\sqrt{2}c_l$.

The choice of the state equation, $\epsilon$$=$$3P$, simplifies the
equilibrium functions as follows,
\begin{eqnarray}{\label{equil1s}}
  f_i^{\rm eq} = w_i n \gamma \left[1+3\frac{(\vec{c}_i \cdot \vec{u})}{c_l^2} \right] \quad ,
\end{eqnarray} 
for $i$$\ge$$0$ and,
\begin{eqnarray}{\label{equil2as}}
  \begin{aligned}
    g_i^{\rm eq} = w_i \epsilon \gamma^2 \biggl[ \frac{1}{\gamma^2
      c_l^2} +4\frac{(\vec{c}_i \cdot \vec{u})}{c_l^2}+
    6\frac{(\vec{c}_i \cdot \vec{u})^2}{c_l^4} -
    2\frac{|\vec{u}|^2}{c_l^2} \biggr] \quad ,
  \end{aligned}
\end{eqnarray} 
for $i$$>$$0$ and,
\begin{eqnarray}{\label{equil2bs}}
  g_0^{\rm eq} = w_0 \epsilon \gamma^2 \left[ 4 - \frac{2+c_l^2}{\gamma^2
      c_l^2} - 2\frac{|\vec{u}|^2}{c_l^2} \right] \quad ,
\end{eqnarray} 
for $i$$=$$0$. Then, the equations for the macroscopic variables take
the form: $n\gamma=\sum_{i=0}^{18} f_{i}$, $\frac{4}{3}\epsilon
\left(\gamma^2 - \frac{1}{4} \right)= \sum_{i=0}^{18} g_{i}^{p}$ and
$\frac{4}{3} \epsilon \gamma^2 \vec{u}=\sum_{i=0}^{18} g_{i}
\vec{c}_{i}$. In our model, the shear viscosity is computed according
to standard LB procedures as: $\eta$$=$$\frac{4}{9}\gamma^2
\epsilon(\tau - \delta t/2)c_l^2$.

\begin{figure}
  \centering
  \includegraphics[scale=0.4]{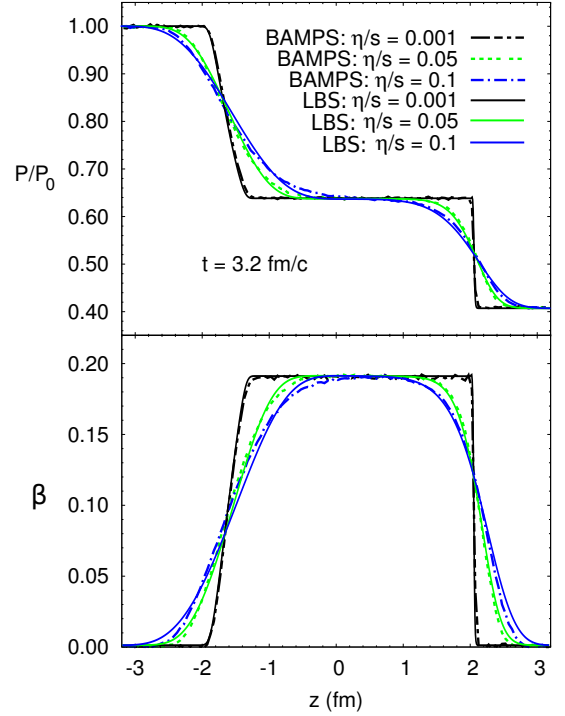}
  \caption{Comparison between the BAMPS simulations\cite{BAMPSs} and
    the lattice Boltzmann results, for $\beta$$\sim$$0.2$. Pressure
    (top) and velocity (bottom) of the fluid as function of the
    spatial coordinate $z$.}\label{compare1}
\end{figure}
\begin{figure}
  \centering
  \includegraphics[scale=0.4]{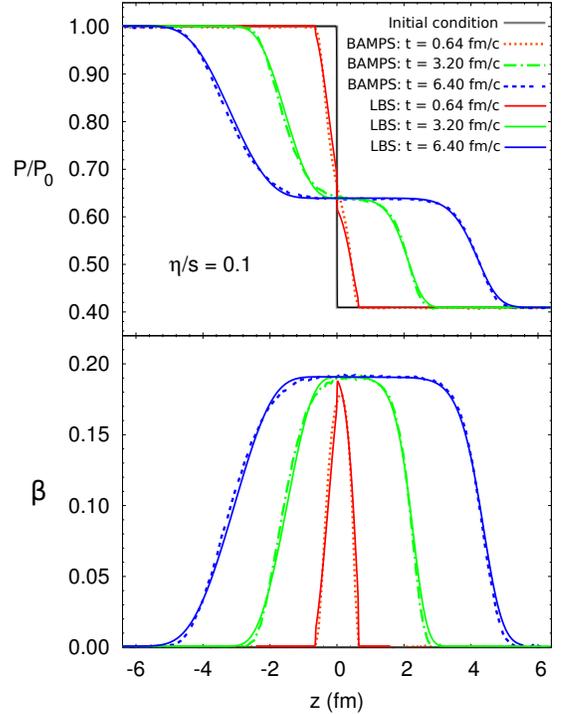}
  \caption{Time evolution of the shock wave for BAMPS
    simulations\cite{BAMPSs} and Lattice Boltzmann results,
    $\beta$$\sim$$0.2$.}\label{comparetime}
\end{figure}

\begin{figure}
  \centering
  \includegraphics[scale=0.5]{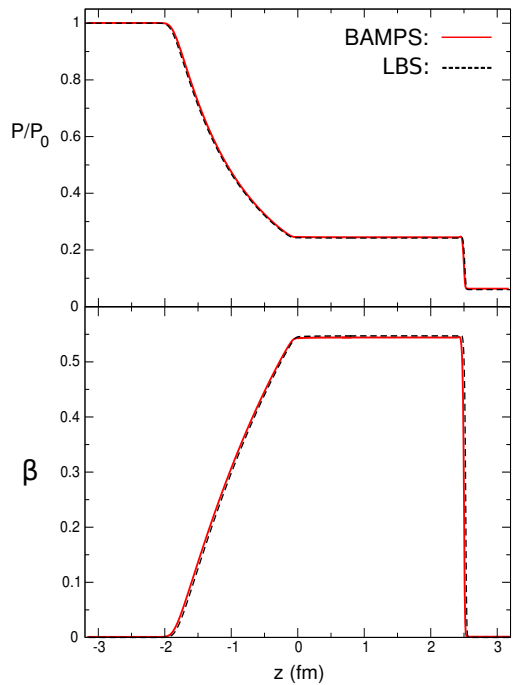}
  \caption{Velocity and pressure profile using numerical ``tachyons''
    at time $t$$=$$3.2$fm/c, with $\beta$$\sim$$0.6$ and
    $\eta/s$$=$$0.001$.}\label{compare3}
\end{figure}

To test the model we solve the Riemann problem in viscous gluon
matter\cite{BAMPSs}. We use the state equation for ultra-relativistic
fluids $\epsilon$$=$$3P$, as before, and the relation between energy
density and particle number density, $\epsilon$$=$$3nT$, with $T$ the
temperature\cite{RelaBoltEqua}. The initial configuration consists of
two regions divided by a membrane located at $z$$=$$0$.  Both regions
have thermodynamically equilibrated matter with different constant
pressure, $P_0$ for $z$$<$$0$ and $P_1$ for $z$$>$$0$. At $t$$=$$0$
the membrane is removed.

We implement a one-dimensional simulation with an array of size
$1$$\times$$1$$\times$$800$. In this case, the 4-velocity is given by
$u^\mu$$=$$(\gamma,0,0,\gamma\beta)^{\mu}$. The velocity of the
lattice is chosen $c_l$$=$$1.0$, therefore the cell size $\delta x$
and time step $\delta t$ are fixed to unity. This corresponds in IS
units to $\delta x$$=$$0.008$fm and $\delta t$$=$$0.008$fm/c. The
viscosity is calculated through $\eta$$=$$\frac{4}{9}\gamma^2
\epsilon(\tau - 1/2)$, and the entropy density by the approximation
$s$$=$$4n-n\ln \lambda$, with $\lambda$$=$$\frac{n}{n^{eq}}$ the gluon
fugacity and the equilibrium particle density $n^{eq}$ are given by,
$n^{eq}$$=$$\frac{d_G T^3}{\pi^2}$ with $d_G$$=$$16$ for gluons. Now,
we can calculate the ratio between the viscosity and entropy density,
$\eta/s$, that is used as a parameter to characterize the shock-wave.
The pressures were chosen $P_0$$=$$5.43$GeVfm$^{-3}$ and
$P_1$$=$$2.22$GeVfm$^{-3}$, corresponding to $7.9433$$\times$$10^{-6}$
and $3.2567$$\times$$10^{-6}$ lattice units, respectively.  The
initial temperature is $T_0$$=$$350$MeV, corresponding to
$T_0$$=$$0.0287$ lattice units. With these parameters, the conversion
between physical and numerical units for the energy, is
$1$MeV$=$$8.2$$\times$$10^{-5}$.

Fig. \ref{compare1} shows the results for different values of $\eta/s$
and the comparison with the BAMPS\cite{BAMPS} (Boltzmann Approach of
Multiparton Scattering) microscopic transport model
simulations\cite{BAMPSs} at time $3.2$fm/c. On the other hand, in Fig.
\ref{comparetime}, we can see the evolution of the system for
$\eta/s$$=$$0.1$ comparing the two numerical models. In both cases, we
find an excellent agreement with BAMPS.  To simulate fluid moving at
higher speed, $\beta$$\sim$$0.6$, we use numerical ``tachyons'' with
$c_l$$=$$10$. Indeed, from Eqs. \ref{equil1} and \ref{equil2a}, it is
seen that the positivity condition $f_i^{\rm eq}$$>$$0$ implies
$\vec{c}_i\cdot\vec{u}$$<$$\frac{c_l^2}{3}$. Now, the pressure $P_1$
is taken as $0.9532$GeVfm$^{-3}$ and we define two temperatures,
$T_0$$=$$0.0328$ and $T_1$$=$$0.0164$, the first one for $z<0$ and the
second one for $z>0$. Fig. \ref{compare3} shows the shockwave for
$\eta/s$$=$$0.001$ and the comparison with the BAMPS
simulation\cite{BAMPSs}, is again excellent.  Our LB scheme easily
extends to three dimensions, as is illustrated in Fig.\ref{picture3D},
where we simulate the collision of a relativistic shock wave with a
fixed spherical obstacle. A typical $200 \times 100 \times 100$
lattice-site simulations spanning $1350$ timesteps, takes about $1900$
CPU seconds on a standard PC.

Summarizing, we have developed a Lattice Boltzmann formulation for
(mildly) relativistic fluids.  One of the major areas of application
of non-relativistic LB schemes is flow through geometrically complex
domains with internal obstacles, such as porous media. It is therefore
expected that the present RLB scheme may become useful for the
simulation of relativistic fluids in complicated geometries.

\subsection{Acknowledgements} 
SS would like to acknowledge kind hospitality and financial support
from ETH Z\"urich.

\bibliography{rlbDec10}

\end{document}